# Experimental investigations on nucleation, bubble growth, and micro-explosion characteristics during the combustion of ethanol/Jet A-1 fuel droplets

D. Chaitanya Kumar Rao[a], S. Syam[a], Srinibas Karmakar[a,†], Ratan Joarder[a]

[a]Department of Aerospace Engineering, Indian Institute of Technology Kharagpur, West Bengal, 721302, India

ABSTRACT

The combustion characteristics of ethanol/Jet A-1 fuel droplets having three different proportions of ethanol (10%, 30%, and 50% by vol.) are investigated in the present study. The large volatility differential between ethanol and Jet A-1 and the nominal immiscibility of the fuels seem to result in combustion characteristics that are rather different from our previous work on butanol/Jet A-1 droplets (miscible blends). Abrupt explosion was facilitated in fuel droplets comprising lower proportions of ethanol (10%), possibly due to insufficient nucleation sites inside the droplet and the partially unmixed fuel mixture. For the fuel droplets containing higher proportions of ethanol (30% and 50%), micro-explosion occurred through homogeneous nucleation, leading to the ejection of secondary droplets and subsequent significant reduction in the overall droplet lifetime. The rate of bubble growth is nearly similar in all the blends of ethanol; however, the evolution of ethanol vapor bubble is significantly faster than that of a vapor bubble in the blends of butanol. The probability of disruptive behavior is considerably higher in ethanol/Jet A-1 blends than that of butanol/Jet A-1 blends. The Sauter mean diameter of the secondary droplets produced from micro-explosion is larger for blends with a higher proportion of ethanol. Both abrupt explosion and micro-explosion create a large-scale distortion of the flame, which surrounds the parent droplet. The secondary droplets generated from abrupt explosion undergo rapid evaporation whereas the secondary droplets from micro-explosion carry their individual flame and evaporate slowly. The growth of vapor bubble was also witnessed in the secondary droplets, which leads to the further breakup of the droplet (puffing/micro-explosion).

# 1. Introduction

Biofuels are widely recognized as the major renewable energy sources which can possibly reduce the dependence on crude oil and can supplement the declining fossil fuels. Jet fuel used in gas turbine engines is one of the primary sources of global carbon dioxide emissions. By overcoming the cost and performance issues, biofuels can be blended with jet fuel or synthetic jet fuels to minimize the emissions [1]. Alcohols as biofuels are being considered as the promising renewable fuels because of their higher volatility and latent heat of vaporization [2-3]. Alcohols, such as ethanol and butanol are already being employed in small proportions with diesel and gasoline in the field of transportation. Complete replacement of jet fuel with ethanol for a gas turbine engine is not practically feasible due to different physico-chemical properties of ethanol [4-6], which may affect the atomization and subsequent combustion processes in the existing gas turbine engines. The addition of ethanol in gasoline, diesel, and biodiesel has been found to reduce engine emissions [7-10]. Blending ethanol with jet fuel could help in reducing soot emissions without significant modifications to the engine fueling and other components of the combustion chamber.

Several experiments on SI and CI engines have been performed by researchers with ethanol as a blend of diesel, biodiesel, and gasoline [7-12]. Droplet combustion experiments have also been conducted on ethanol and their blends with gasoline and diesel [13-16]. It is well understood that burning the mixtures of alcohol/alkane results in disruptive behavior due to significant volatility differential between the fuel components [17]. However, due to the small volatility differential between ethanol and gasoline, the combustion of their blends does not result in disruptive behavior. Similarly, in spite of the significant volatility differential between ethanol and diesel, ethanol is completely immiscible in diesel. Ivanov et al. [18] first reported the micro-explosion phenomenon, and it was demonstrated that water in emulsified fuels improves the combustion process due to the breakup of droplets. It has been widely acknowledged that the volatility differential among fuel constituents in a high-temperature environment results in formation of bubble embryo inside the parent droplet. The subsequent growth of this bubble leads to breakup of the droplet. The disintegration of multi-component droplet due to this significant volatility differential is known as 'micro-explosion' or secondary atomization. In practice, the occurrence of micro-explosions has the potential to enhance fuel atomization in the combustion chamber, which is considered as an effective way of promoting efficient combustion in an engine. Several numerical and experimental studies have been performed to understand the micro-explosion behavior of emulsified fuel droplets [19-28]. The overall benefits of the disruptive burning in emulsions has also been confirmed and visualized in spray flow and engines [27-28].

The micro-explosion phenomenon has been widely studied in miscible fuel droplets in normal gravity as well as in freely falling conditions [29-36]. Lasheras et al. [29] and Wang et al. [17,31] performed a series of experiments on free droplets of alcohol/alkane solutions and emulsions. It is well established from experiments that disruptive burning is primarily dependent on the difference in boiling points of the lower and higher volatile fuel constituents as well as on the appropriate range of relative concentration of alcohol and alkane constituents. Zeng et al. [35] and Shen et al. [36] developed numerical models to characterize the micro-explosion phenomenon for bio-fuel droplets. The proposed model was used to determine the Sauter mean radius (SMR) and velocity of secondary droplets from micro-explosions. The numerical model was also used to characterize the onset of micro-explosion for binary and tertiary fuel droplets. More recently, in our previous work, an experimental investigation was performed to study the puffing and micro-explosion events in butanol/Jet A-1 and A-B-E/Jet A-1 fuel droplets [37]. The onset of micro-explosion was characterized by normalized squared onset diameter (NOD), and it was found that NOD is almost similar for butanol blends and A-B-E blends. The fuel blends with comparable volatility differential and a similar proportion resulted in nearly identical bubble growth rates and Sauter mean diameters of secondary droplets.

Although several studies have been conducted on ethanol/gasoline and ethanol/diesel droplets, an experimental investigation on the combustion characteristics and disruptive behavior of ethanol/Jet A-1 droplets is lacking. The focus of the present work is to understand and compare the nucleation, bubble growth and subsequent breakup among different blends of ethanol/Jet A-1. An attempt has been made to differentiate and compare the disruptive characteristics of partially miscible blends (ethanol/Jet A-1) with our previous work on completely miscible blends (butanol/Jet A-1).

## 2. Experimental Methodology

The single droplet experiments were performed under quiescent atmospheric conditions (25 ± 5 °C, 1 atm, and RH ~75 ± 5%) in a cylindrical stainless steel chamber of dimensions 500 mm x 200 mm. The chamber consists of two quartz windows for optical access and backlighting. A schematic diagram of the experimental setup is illustrated in Fig. 1. A micro-pipette was used to produce constant volume droplets of 2 ± 0.05 μl, which were suspended on a 0.2 mm diameter quartz fiber. The quartz fiber was utilized in the experiments because of its low thermal conductivity (1.4 W/m K). The equivalent diameter of the suspended fuel droplets is 1.7 ± 0.1 mm.

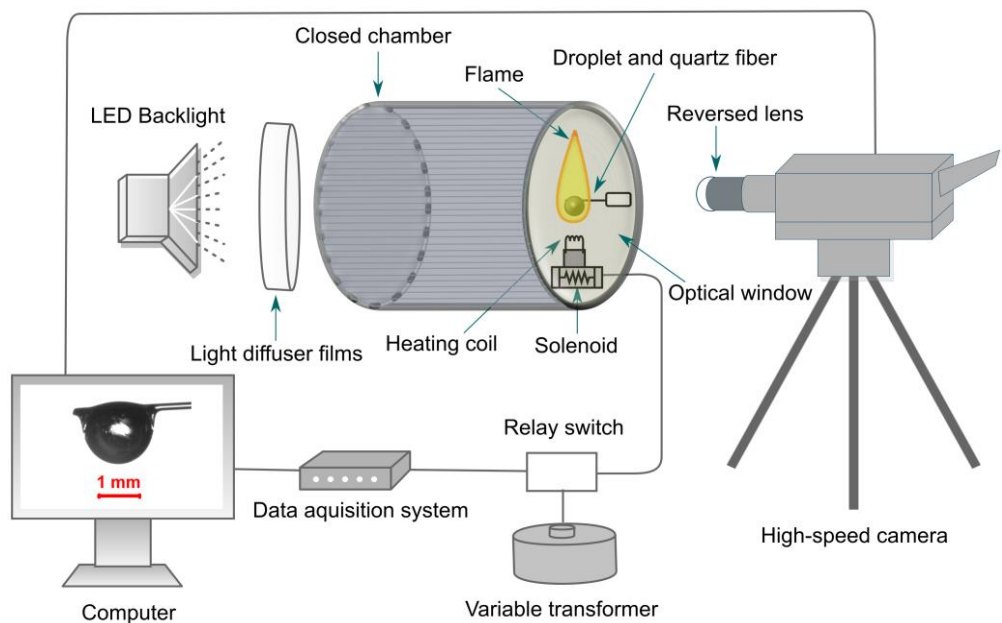

**Fig. 1.** Schematic of the experimental apparatus.

The fuels studied in the present work are Jet A-1 and ethanol; whose selected properties are listed in Table 1. Three combinations of ethanol/Jet A-1 are considered in the present study. The composition of fuel blends constituting a test droplet is shown in Table 2. The fuel blends were prepared using an ultra-sonicator; that was operated at 250 watts for 20 seconds to ensure proper mixing. The fuel droplets were ignited by a 15-volt DC supply through a coiled nichrome wire of 0.5 mm diameter, which is held by a solenoid. The solenoid moves the coil beneath the droplet, and during ignition, the coil is located at a distance of 1 ± 0.2 mm from the bottom surface of the droplet. The nichrome wire supplies heat (470 ± 15 °C) to the droplet and it is retracted via the data acquisition system (DAQ) as soon as the heating time is ended. The retraction velocity of the nichrome coil is maintained at around 0.15 m/s such that it does not create any disturbance to the droplet flame. Therefore, the interaction of nichrome wire with the flame is minimal, and the wire is assumed not to play a significant role in the evolution of droplet. True color images of the droplet burning sequence were captured using a DSLR camera to differentiate the visual appearance of the flame structure and sooting tendencies of different fuel blends. A high-speed monochrome camera (Phantom v7.3) was used to record the droplet burning process at a resolution of 800 x 600 pixels at 3000 fps. A multi-LED backlight consisting of 24 high power LEDs was used to illuminate the fuel droplets. In order to capture the high-speed flame images, backlighting intensity and camera exposure were optimized to visualize both the droplet flame and the growth of vapor bubble inside the droplet. Proper care was taken to preserve the backlighting intensity, camera exposure, and other experimental conditions identical. This also permits a valid and efficient basis of assessment of sooting propensities among the captured photographs. A gated intensified CCD camera (4 Quick E, Stanford Computer Optics, Inc) with a maximum resolution of 1360 x 1024 pixels was used to capture chemiluminescence of the electronically excited methylidyne radical (CH*) during the temporal evolution of the droplet flame. An in-house MATLAB code was employed to compute the equivalent diameter of the droplet as a function of time. The fiber is used as a reference scale to obtain the scale factor (for pixels to mm conversion). The equivalent diameter of the droplet was calculated using the relation, $D = \sqrt{(D_h D_v)}$, where $D_h$ and $D_v$ are the major and minor axes of the ellipse (droplet). An image analysis platform, Image-Pro Plus (version 6.0) from Media Cybernetics, was used to determine the growth of vapor bubble, and the diameter and velocity of secondary droplets. To obtain the diameter and velocity of secondary droplet, the image sequence is first processed by adjusting the threshold such that the background is completely dark and the object of interest (droplet) is white. The software then automatically identifies the boundary of the secondary droplet and calculates its mean diameter and velocity. The bubble diameter is determined using a manual approach by placing a virtual circle on the image and positioning it based on personal judgement of the bubble boundary. The ambiguity in the measurement of bubble diameter is ± 0.05 mm, which arises mainly due to the asymmetric shape of the bubble during its initial growth. Similarly, the uncertainty in the measurement of the ejected droplet diameter following the breakup of parent droplet is ±10 μm. To verify the repeatability and to carry out a probabilistic analysis of obtained results, the experiments were performed 25 times for each blend case.

**Table 1**
Properties of the fuels investigated in this study.

| Physical Properties | Jet A-1[a] (Standard) | Ethanol[b] | Butanol[b] |
|---|---|---|---|
| Molecular Formula | $C_8$-$C_{16}$ | $C_2H_5OH$ | $C_4H_{10}O$ |
| Boiling point (°C) | 180-250 | 78.4 | 117.7 |
| Reid vapor pressure (kPa) | < 1 | 16 | 2.2 |
| Density at 15 °C (kg/m$^3$) | 775–840 | 795 | 813 |

[a]Standard specifications of Jet A-1 are from ASTM D1655, [b]Properties of ethanol and butanol are from Ref. [4-6].

**Table 2**
Composition and nomenclature of fuel blends.

| Composition of fuel mixture (volume basis) | Designated nomenclature |
|---|---|
| 10% ethanol, 90% Jet A-1 | E10 |
| 30% ethanol, 70% Jet A-1 | E30 |
| 50% ethanol, 50% Jet A-1 | E50 |

## 3. Results and Discussion

### 3.1 Flame appearance and sooting propensity

The flame images corresponding to Jet A-1, pure ethanol, and E50 droplets show a characteristic envelope flame surrounding the droplets (Fig. 2). Pure Jet A-1 droplets burn with bright yellowish flame indicating high sooting tendency. In contrast, pure ethanol droplets burn with a light bluish flame and a relatively spherical flame structure. The different flame shapes of Jet A-1 and ethanol droplets is due to the variation in the natural convective heat transfer, density, and sooting propensity of the fuels. The lower sooting propensity of ethanol is attributed to the presence of oxygen atom in its molecular structure. Apparently, the sooting tendency decreases in the order of Jet A-1, E50, and pure ethanol. The flame images indicate smooth burning of jet fuel and pure ethanol as seen in Fig. 2 (i)-(ii). Due to large volatility differential among the fuel components, E50 droplets showed disruptive nature (micro-explosion) as evident from perturbation of flame in the sequence of color images (Fig. 2 (iii)). Ejected blue flame in the photographic images indicates burning of expelled ethanol droplets during the disruptive events.

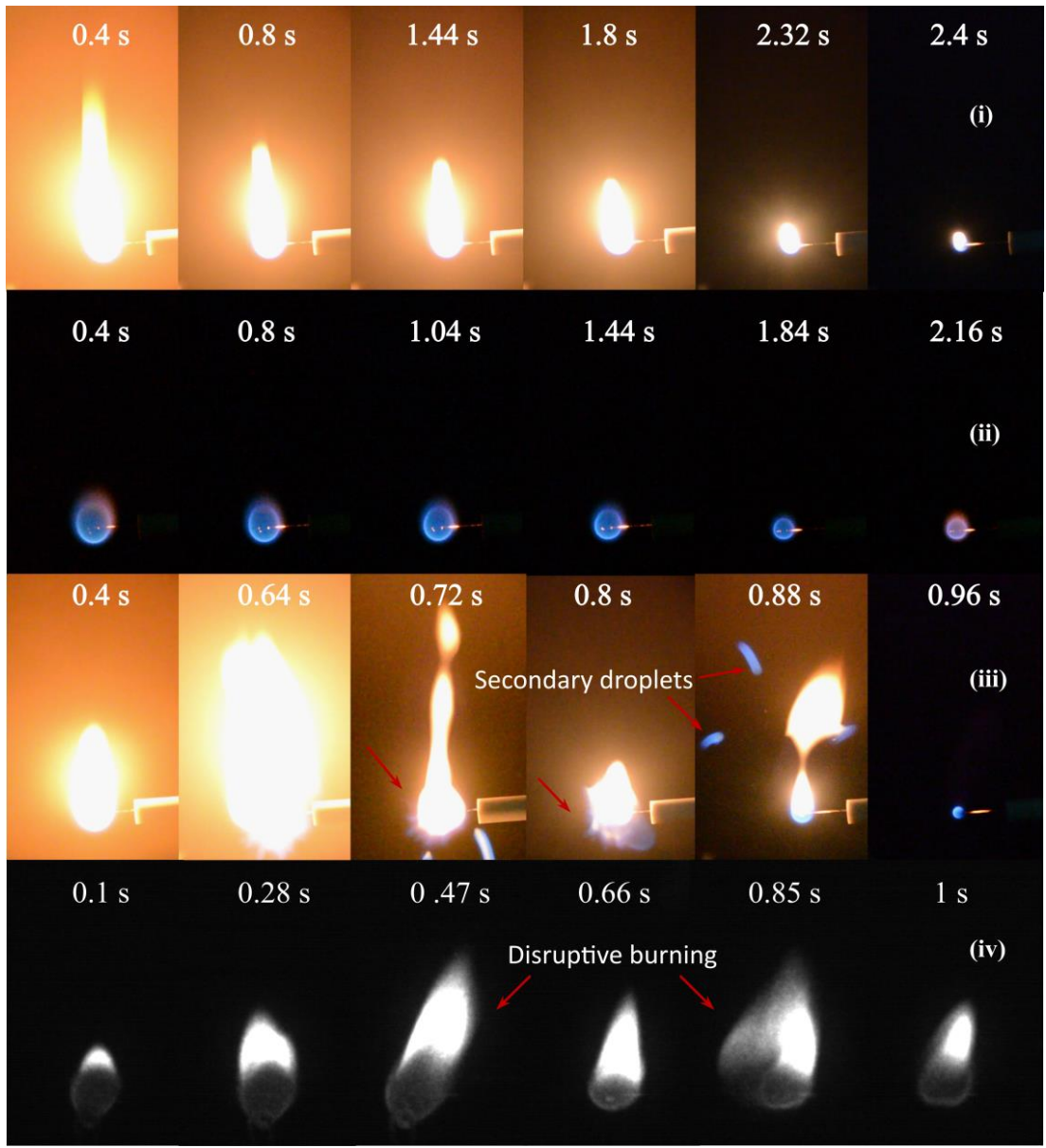

**Fig. 2.** The sequence of flame images representing smooth burning of (i) Pure Jet A-1 and (ii) pure ethanol, and disruptive burning of (iii) E50 droplet and (iv) CH* chemiluminescence images of burning E50 droplet. Arrows indicate disruptive behavior due to higher volatility differential.

The shift in the flame orientation due the disruptive behaviour is also evident from the CH* chemiluminescence images shown in Fig. 2 (iv). Due to the low exposure time capability of the camera, the background noise is significantly reduced, and weak intensities of chemiluminescence are captured effectively. The orientation of the perturbed flame indicates the ejection of secondary droplets in the direction opposite to that of disturbed flame. In particular, the flame disruption suggests the expulsion of secondary droplet/ethanol vapor subsequent to the breakup of a vapor bubble. The ejected vapor and secondary droplet ignites at the flame front thereby increasing the chemiluminescence intensity.

*3.2 Nucleation and bubble growth*

Homogeneous nucleation was observed in all the blends of ethanol/Jet A-1, which is in disparity with butanol/Jet A-1 blends, where the nucleation was not favored in 10% butanol blend. The nucleation was probable in E10 due to increased volatility differential in spite of having a lower proportion of ethanol. The onset of bubble nucleation is characterized by normalized squared onset diameter (NOD), which is defined as the square of the ratio of droplet diameter at the instant of nucleation to initial droplet diameter. The NOD was obtained ostensibly from the high-speed images. The most probable NOD values for E10, E30, and E50 are 0.8, 0.9, and 0.9 respectively (Fig. 3). The NOD for ethanol/Jet A-1 droplets is higher than that of butanol/Jet A-1 blends since the volatility difference of ethanol/Jet A-1 combination is greater compared to that of butanol/Jet A-1. The larger NOD of ethanol blends implies relatively earlier nucleation of vapor bubble due to the lower superheat limit [38] of ethanol (~466 K) compared to butanol (~512 K). This nature is consistent with the computational results reported by Shen et al. [36].

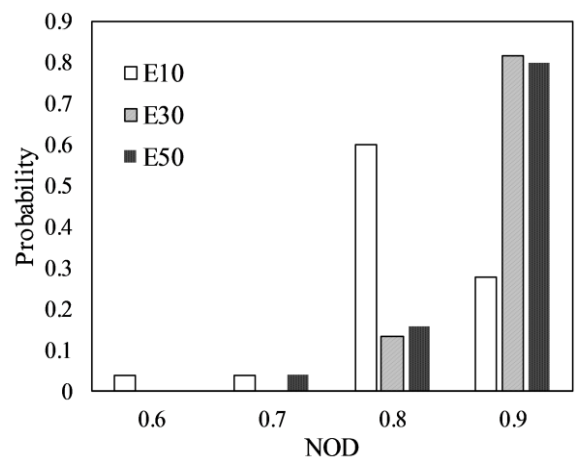

**Fig. 3.** The probability of NOD values for the blends of ethanol.

The vapor bubble growth inside ethanol/Jet A-1 droplets is shown in Fig. 4. The bubble growth for butanol/Jet A-1 droplets is also plotted for comparison. The bubble growth rates corresponding to the most likely state of nucleation for E30 and E50 fuel droplets are nearly the same. However, the rate of bubble growth is faster in the case of ethanol blends compared to that of butanol blends. This rapid expansion might be due to the larger volatility differential among the components of ethanol blends. It is also evident that the pre-breakup bubble diameter increases with increase in the proportion of higher volatile component. This is a consequence of increased nucleation sites and the subsequent coalescence of bubbles leading to the formation of a bigger bubble.

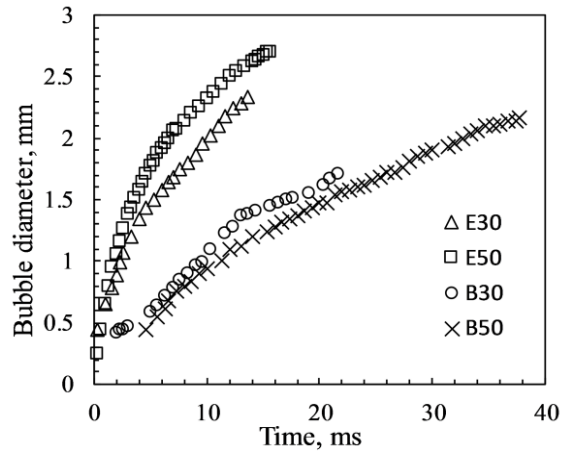

**Fig. 4.** Temporal evolution of bubble diameter for different blends. B30 and B50 correspond to 30% and 50% butanol blends.

*3.3 Photographic sequences and regression profiles of burning droplets*

The disruptive events in various fuel blends are represented by photographic sequences of the combustion process as well as through the evolution of droplet diameters. A schematic diagram of puffing, micro-explosion, and abrupt explosion is illustrated in Fig. 5 (a)-(c).

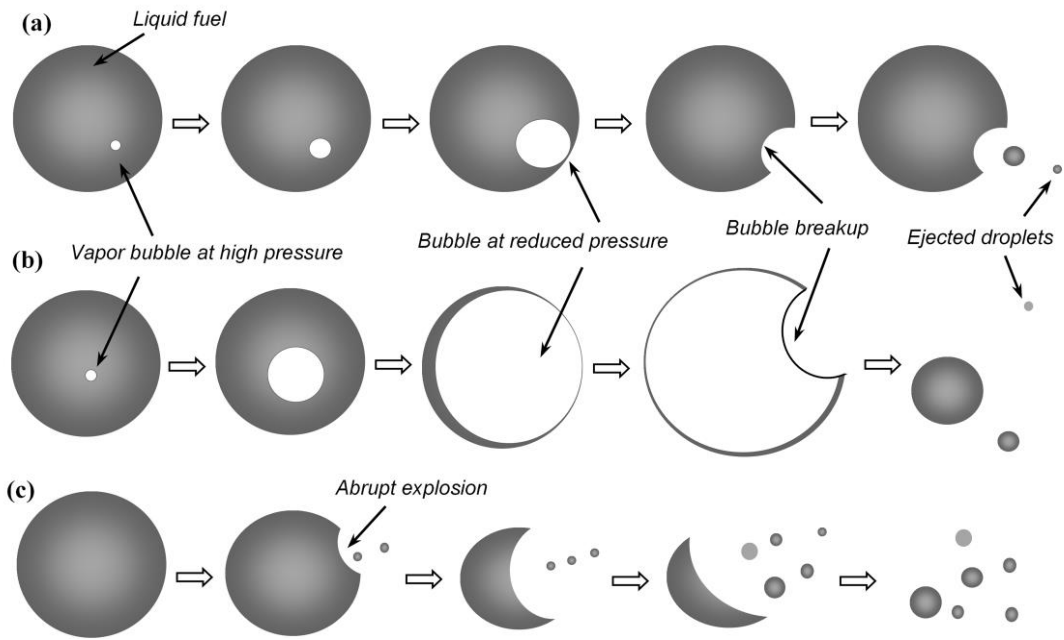

**Fig. 5.** Schematic diagram of (a) puffing, (b) micro-explosion, and (c) abrupt explosion. The sequence represents nucleation, bubble growth and the breakup of parent droplet.

Due to the lower proportion of higher volatile component, fewer nucleation sites are formed, which leads to the formation of a moderately small bubble. When this bubble does not grow further, breakup occurs resulting in the ebullition of fine droplets. This characteristic of droplet ejections is referred as puffing. A larger fraction of higher volatile component leads to the generation of a significant number of nucleation sites, which results in the formation of a bigger bubble. The consequent breakup of this bubble results in micro-explosion. In addition to puffing and micro-explosion, the abrupt explosion was also observed in the present work. As discussed before, puffing and micro-explosion occur due to the collapse of a bubble, whereas an abrupt explosion takes place without any noticeable bubble inside the droplet and leads to complete disintegration of parent droplet. Since this phenomenon was not observed in our previous work [37] where butanol and A-B-E were blended with Jet A-1, it seems that the sudden explosion arises due to the limited miscibility of the blends. It can be inferred that due to limited nucleation sites, the microscopic bubbles do not coalesce to grow further. Instead, the high-pressure bubble completely shatters the droplet releasing the immense pressure within it. Apart from the volatility differential required to generate the vapor bubble, the limited miscibility of ethanol in Jet A-1 [39] can be conjectured to play a major role here since the immiscibility adversely affects the mobility of microscopic bubbles. Similar characteristics of the sudden explosions (without any visible bubble breakup) have been reported during the combustion of emulsion fuels such as water-in-oil emulsions and diesel-biodiesel-ethanol blends [25,40-41].

The nucleation, bubble growth and resulting fragmentation of droplet by the disruptive events are presented through a sequence of high-speed images. Only a few frames of foremost prominence are carefully chosen and presented here.

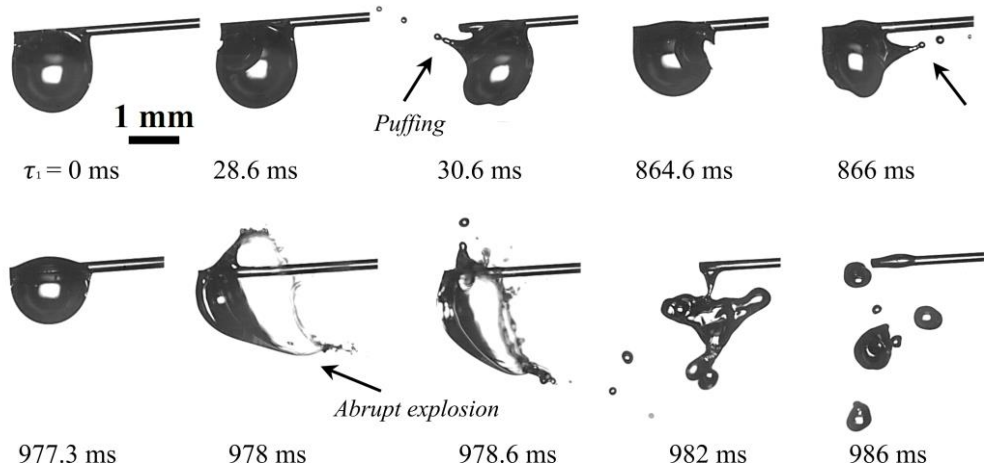

**Fig. 6.** The sequence of images of disruptive events in E10 blend.

Figure 6 accounts for a typical image sequence of combustion of an E10 droplet, where $\tau_1 = 0$ ms signifies the onset of homogeneous nucleation. Soon after the nucleation, puffing occurs at different time intervals. Even though the volatility differential in E10 blend is sufficient for the occurrence of nucleation, the fraction of ethanol is inadequate for the bubble to grow significantly. Once the droplet becomes steady after a series of puffing events, it experiences abrupt explosion resulting in the complete shattering of the droplet. As seen in the figure, prior to the abrupt explosion, there is no indication of the presence of vapor bubble inside parent droplet. Since ethanol is not completely miscible in Jet A-1 and also due to its lower proportion in E10, the microscopic bubble does not grow further, which leads to abrupt explosion. Therefore, the abrupt explosion is dominant in this particular blend, and the probability of abrupt explosion was found to be around 76%. As the percentage of ethanol is increased, nucleation sites inside the droplet also rise which favors the bubble growth and further leads to micro-explosion. Figure 7 (a) represents a typical burning sequence of E30 droplet where the vapor bubble grows and collapses repetitively. The disruption created by puffing generates a turbulence inside the droplet which in turn supports the formation of more nucleation sites [26,37,42]. Since the presence of a vast number of nucleation sites leads to the formation of a bigger bubble, the vapor bubble breaks apart at 320 ms. Combustion process endures even after the micro-explosion event whereas it finishes immediately after the fragmentation by abrupt explosion.

Though this sequence (continuous puffing followed by micro-explosion) is the most prospective occurrence (68% probability) in E30 blend, the abrupt explosion was also observed (36% probability).

A typical sequence of bubble growth leading to the breakup of E50 droplet is presented in Fig. 7 (b). The vapor bubble that has grown from the nucleus can be seen to be positioned near the center of the droplet at $\tau_1 = 0.66$ ms. The scattering of light highlights the bubble periphery inside the droplet, and bright light at the center is due to the optical effects. The vapor bubble grows symmetrically throughout its lifetime until it ruptures at 16.6 ms, resulting in the ejection of secondary droplets. This nature is dissimilar to the bubble growth in butanol blends where the heavier liquid (Jet A-1) distorts the sphericity of parent droplet. The symmetrical growth of vapor bubbles in ethanol blends can be attributed to the comparatively faster growth rate of the bubble.

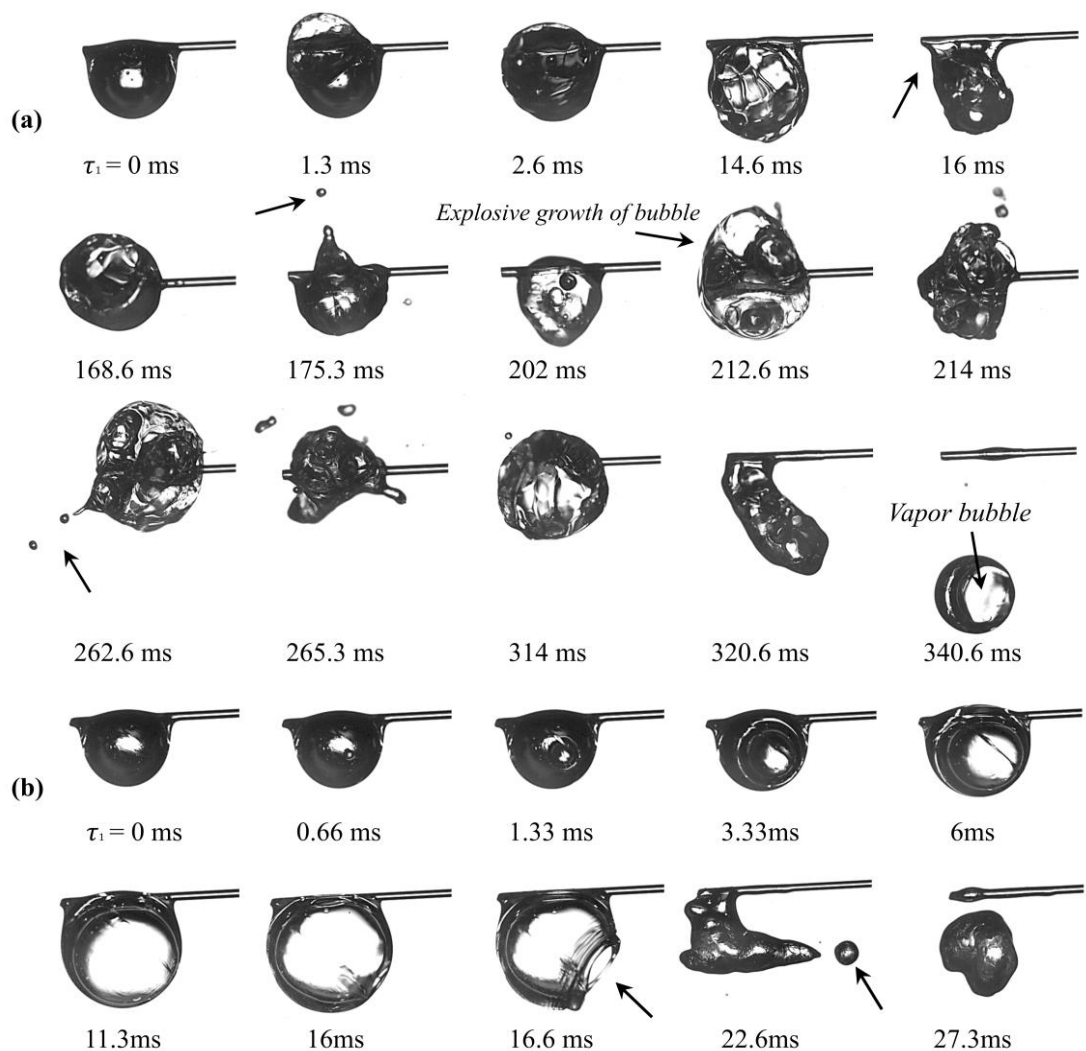

**Fig. 7.** Bubble growth and disruptive behavior associated with the most likely state of nucleation of (a) E30 and (b) E50 blends. The arrows indicate breakup of parent droplet and expulsion of secondary droplets.

The probability of micro-explosion is moderately higher (more than 90%) for E50 droplets than that for E30 droplets. Similarly, the possibility of micro-explosion in ethanol blends is considerably greater compared to butanol blends. It is evident that both volatility differential and proportion of higher volatile component are playing a significant part in the occurrence of micro-explosion. It can also be contemplated from Fig. 7 that as the percentage of the higher volatile component is raised, the breakup intensity as well as the probability of droplet breakup increase. This high-intensity breakup can be credited to the rapid growth rate of vapor bubble and relatively larger size of the inflated droplet [29].

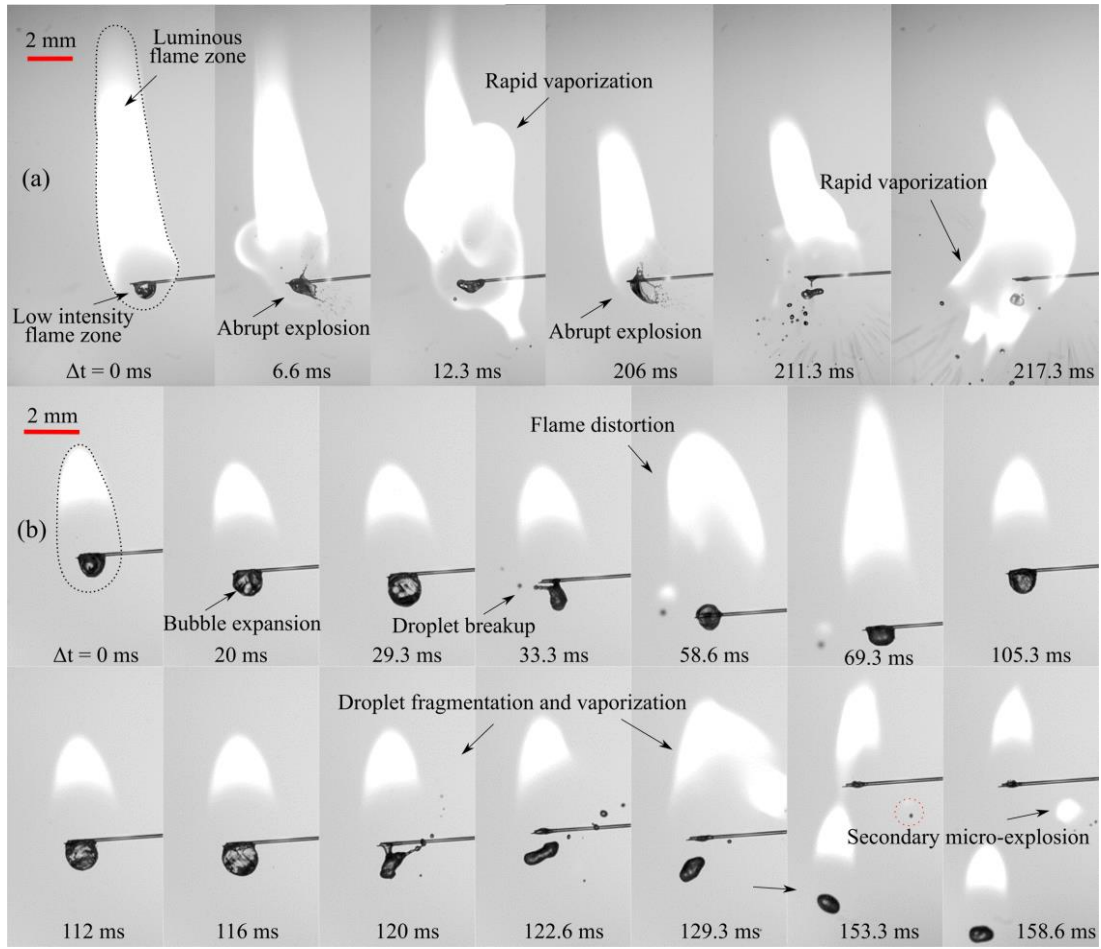

**Fig. 8.** Large-scale flame distortion following the breakup of parent droplet due to (a) Abrupt explosion (b) Micro-explosion.

The breakup of parent droplet due to abrupt explosion/micro-explosion significantly distorts the droplet flame. Figure 8 (a) shows the high-speed flame images corresponding to a typical abrupt explosion event. The envelope flame exhibits two distinct zones, i.e., luminous flame zone and low-intensity flame zone. The backlight, which is employed for simultaneous visualization of the droplet, vapor bubble, and, flame causes the low-intensity flame zone to appear almost invisible. As seen in the figure, soon after the abrupt explosion ($\Delta t$=6.6 ms), the secondary droplets are transported to the flame. The fine secondary droplets are then engulfed in the flame resulting in rapid vaporization (Δt=12.3 ms). The parent droplet again undergoes abrupt explosion (Δt=206 ms), where the complete disintegration of the droplet creates large-scale flame distortion and instant vaporization of the ejected droplets. Similarly, Fig. 8 (b) represents the sequence of images of flame distortion caused by micro-explosion. Since the diameter of the secondary droplets produced from micro-explosion is relatively larger than that of the abrupt explosion, the droplets carry their own flame after moving out of the envelope flame. Therefore, the evaporation of secondary droplets is not as rapid as that of the abrupt explosion.

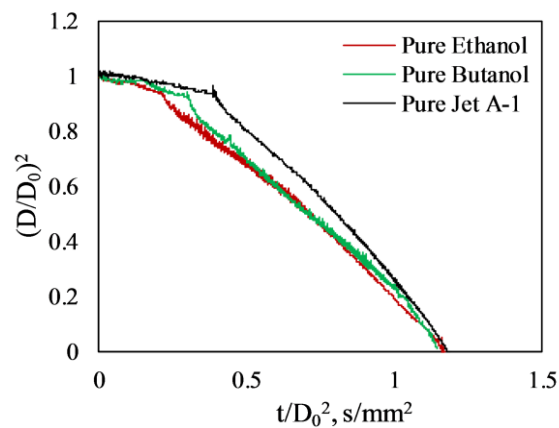

**Fig. 9.** The temporal evolution of droplet diameter of pure ethanol, pure butanol, and Jet A-1 droplets.

Figure 9 shows a comparison of the temporal evolution of droplet diameters for pure ethanol, pure butanol, and Jet A-1 following the $D^2$ regression rate law. The curves indicate a smooth and uninterrupted evaporation of droplets without any disruptive events. In contrast, Fig. 10 represents the disruptive nature of droplets during the evolution of droplet diameter of ethanol/Jet A-1 blends that relate to the sequence of images shown in Figs. 6 and 7.

As seen in Fig. 10 (a), the initial sudden rise in droplet diameter followed by its immediate rapid drop is due to the expansion and subsequent collapse of the bubble. The first spike corresponds to the bubble growth leading to puffing while the final peak corresponds to the abrupt explosion of parent droplet. The highest peak in Fig. 10 (b)-(c) represents the maximum droplet diameter where the bubble ruptures leading to micro-explosion. The time period from bubble generation to the breakup is extremely short (order of 1/100 of average droplet lifetime).

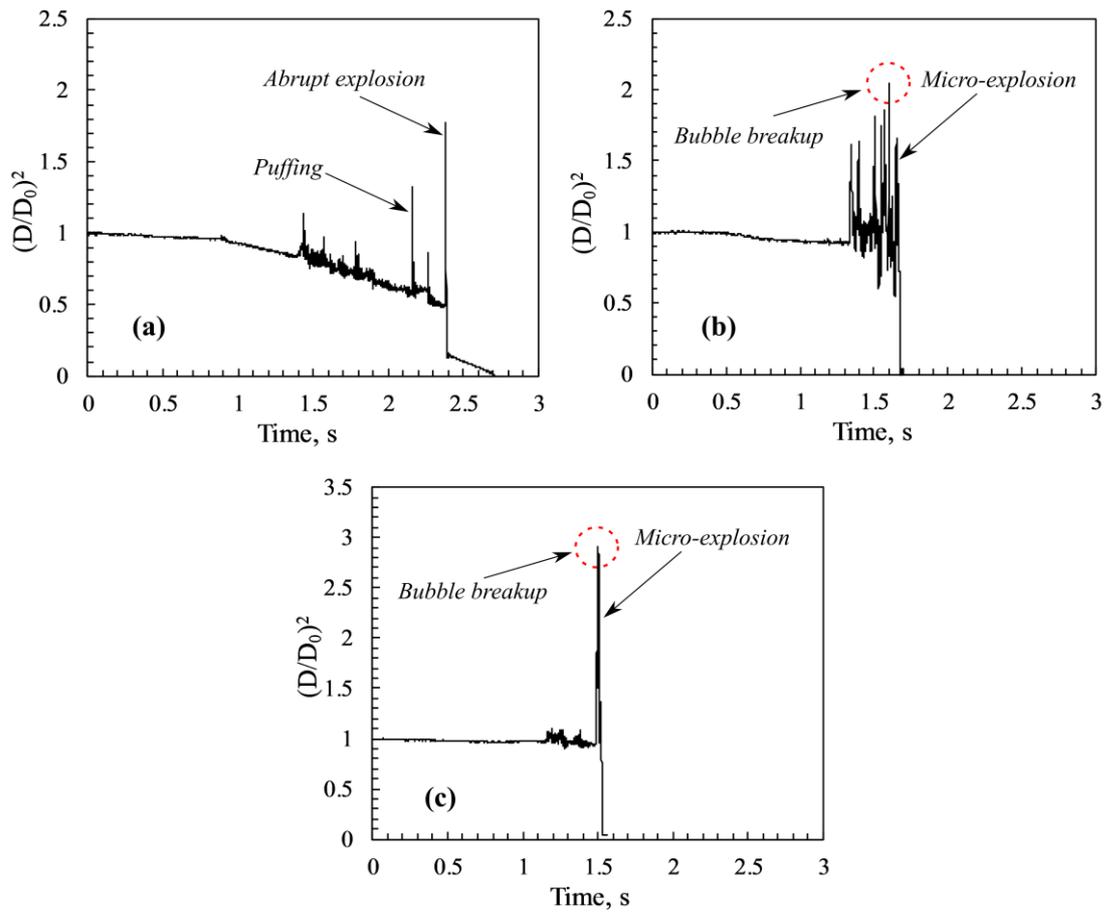

**Fig. 10.** Temporal evolution of droplet diameter for a) E10, b) E30, and c) E50 blends associated with most probable NOD.

The characteristic features of the disruptive burning in ethanol/Jet A-1 blends are highlighted in Table 3. Similar to the combustion characteristics of butanol/Jet A-1 blends [37], an increase in the proportion of ethanol enhances the probability of micro-explosion and hence reduces the average droplet lifetime in ethanol/Jet A-1 blends. It is also noticed that the average droplet lifetime of ethanol/Jet A-1 blends is appreciably lower than that of butanol/Jet A-1 blends for the same blending proportion.

**Table 3**
Characteristics features of disruptive burning in different blends.

| Blends | Dominant characteristics of burning | Probability of Abrupt explosion | Probability of Micro-explosion | Avg. droplet lifetime (relative to pure Jet A-1) |
|---|---|---|---|---|
| E10 | Abrupt explosion | 76% | - | 60% |
| E30 | Micro-explosion | 36% | 68% | 43% |
| E50 | Micro-explosion | 4% | 84% | 40% |

The probability of the main disruptive phenomena, i.e., micro-explosion and abrupt explosion are represented with the variation of the concentration of the volatile component in Fig. 11. It is evident from the figure that the probability of abrupt explosion reduces with the increase in the proportion of the volatile component in ethanol/Jet A-1 blends. The increase in the percentage of volatile component favors higher nucleation sites and hence the coalescence of bubbles, leading to micro-explosion. For small proportions (like E10), immiscible volatile fuel (ethanol) stays in packets within the droplet and bursts. For similar small proportions, miscible blend (butanol/Jet A-1) remains mixed (distributed) in the entire droplet and hence does not exhibit abrupt explosion. At a little higher concentration of volatile component (ethanol) in Jet A-1, the volatile component stays as undissolved packets as well as dissolved state for ethanol leading to both abrupt explosion (less probable) and micro-explosion (more probable). In the case of butanol, it exists only in the dissolved state and hence results in only micro-explosion. In general, the probability of disruptive nature is higher for ethanol/Jet A-1 blends compared to butanol/Jet A-1 blends.

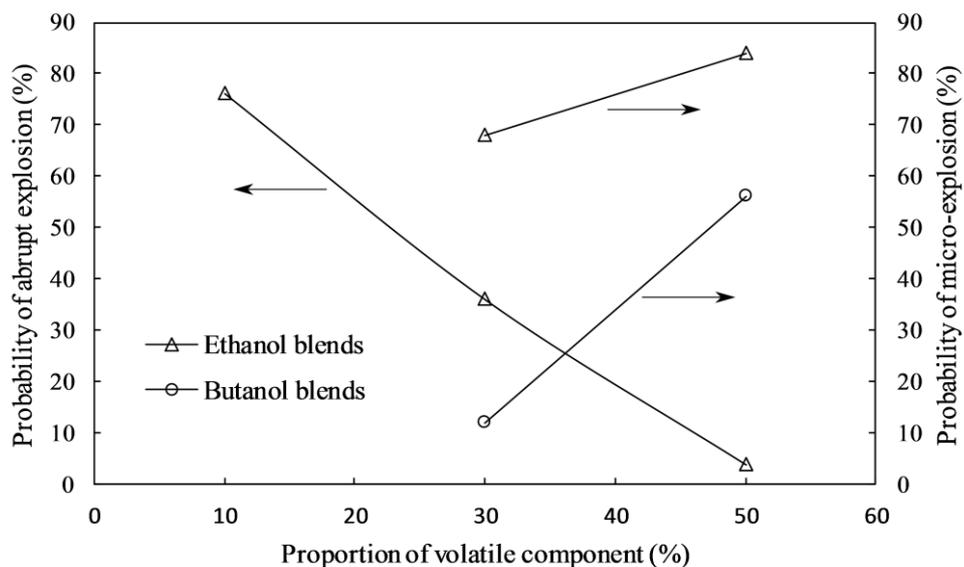

**Fig. 11.** Probability of abrupt explosion and micro-explosion in ethanol/Jet A-1 and butanol/Jet A-1 blends.

*3.4 Breakup characteristics of fuel droplets*

Diameter versus velocity distribution of secondary droplets resulting from the abrupt explosion, puffing, and micro-explosion are shown in Fig. 12.

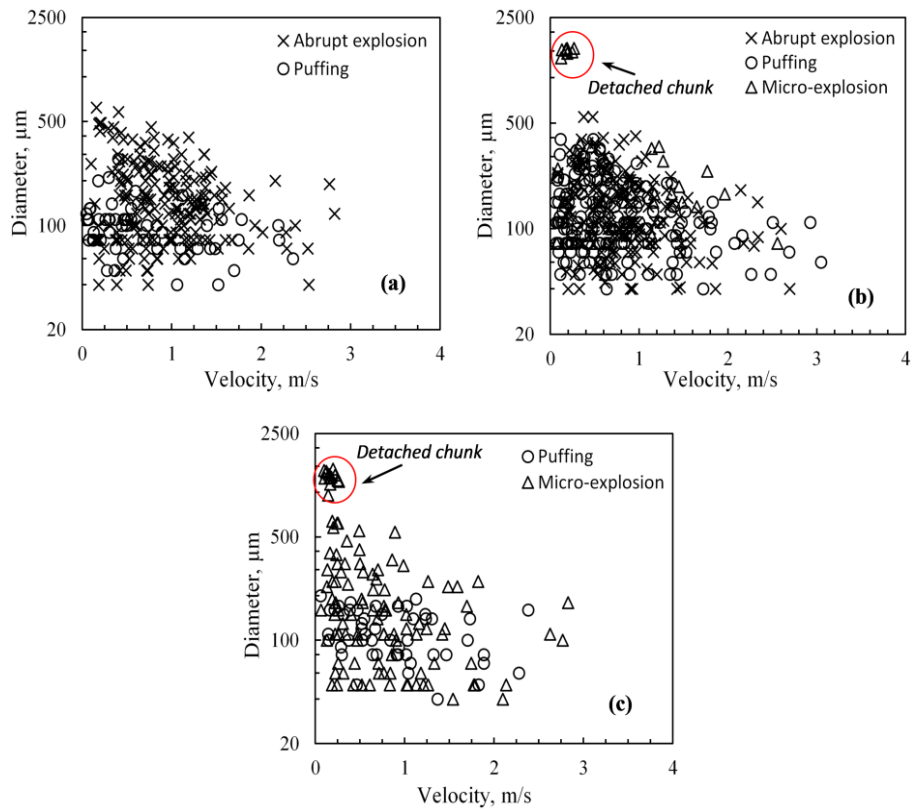

**Fig. 12.** Diameter vs. velocity distribution of secondary droplets generated by the abrupt explosion, puffing, and micro-explosions in a) E10, b) E30, and c) E50 droplets. The circles indicate the diameter of the detached chunk after micro-explosion.

As the secondary droplets are scattered during the disruptive events, not all droplets are in-focus (or in the same plane) in the images. The secondary droplets that were in-focus are considered while the out of focus droplets are ignored for accuracy. The diameter of secondary droplets is presented in the logarithmic scale to provide a clear insight into the diameter distribution. Figure 12 (a) shows the diameter and velocity distribution of E10 droplets where the abrupt explosion is the dominant phenomenon. Abrupt explosion leads to the expulsion of multiple droplets with both larger and smaller diameters. Puffing seems to produce relatively smaller diameter droplets compared to micro-explosion as seen in Fig. 12 (b)-(c). Since micro-explosion was the dominant event in E30 and E50 blends, a detached chunk of significant diameter (900-1500 μm) was observed. The Sauter mean diameter (SMD) of ejected droplets (excluding the detached lump) is 305, 300, and 390 μm for E10, E30, and E50 respectively. The SMD of the secondary droplets for ethanol blends is approximately 20% of the initial droplet diameter. The velocity of ejected secondary droplets varies from around 0.05 m/s to 3 m/s. The smaller diameter secondary droplets were observed to possess higher velocity.

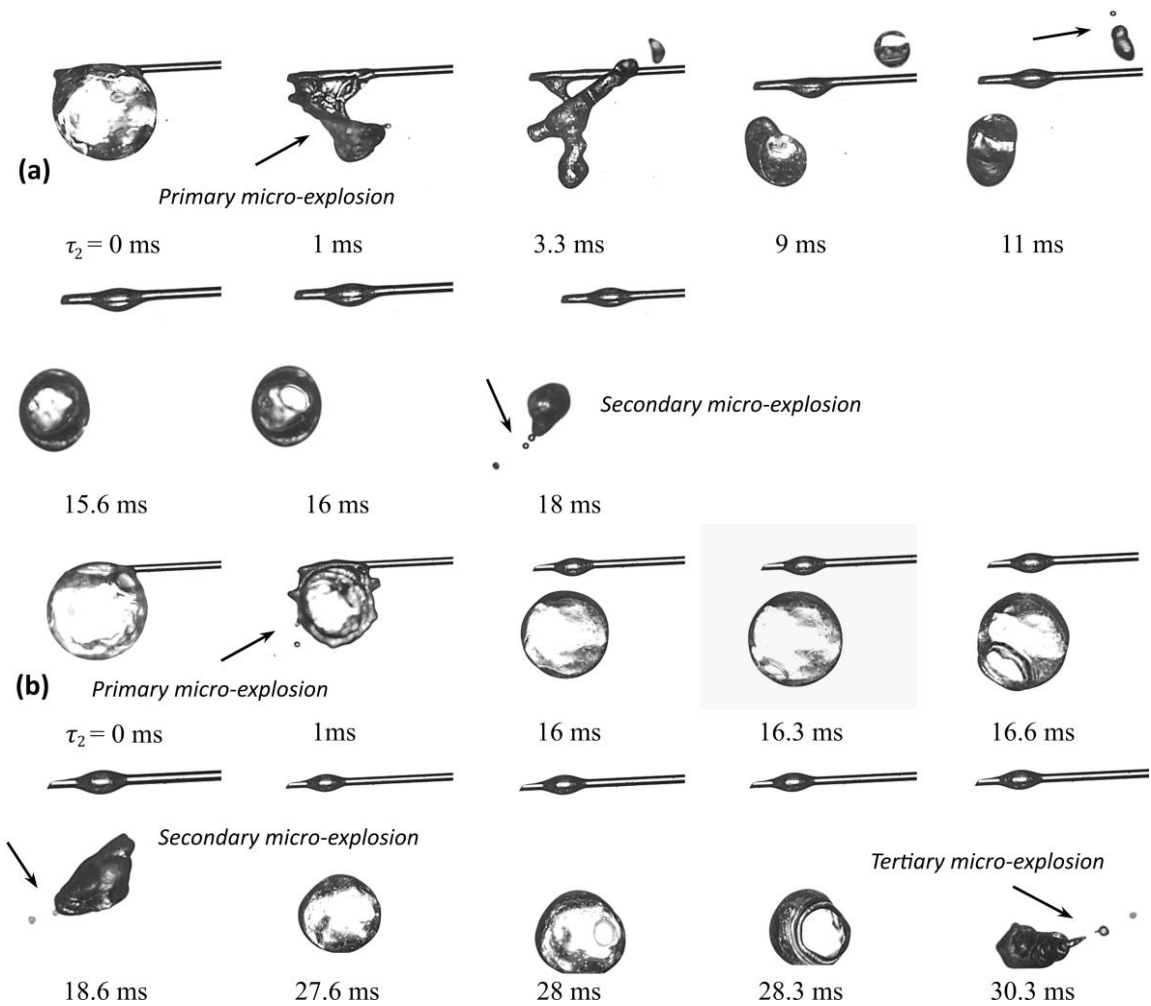

**Fig. 13.** Puffing/micro-explosion events in secondary and tertiary droplets.

Bubble growth in secondary droplets and the subsequent puffing/micro-explosions are also captured in the present work. Some sequence of images of multiple puffing/micro-explosions in E50 is shown in Fig. 13 (a)-(b). Here, $\tau_2 = 0$ ms denotes the onset of primary micro-explosion in the parent droplet. As seen in Fig. 13 (a), after the occurrence of primary puffing or micro-explosion, a bubble starts to grow in the detached droplet and consequently breaks apart (secondary micro-explosion) leading to the ebullition of fine droplets. The diameter of ejected droplets from the secondary micro-explosion is in the range of 100 to 300 microns. Interestingly, tertiary micro-explosion was also observed in the tertiary droplets due to the presence of a considerable amount of higher volatile component (Fig. 13 (b)). This indicates that the micro-explosion events can continuously occur until the proportion of higher volatile component is inadequate for further bubble growth in the droplets.

## 5. Conclusions

An experimental investigation is carried out on droplet combustion of ethanol/Jet A-1, and the obtained results were compared to that of butanol/Jet A-1 blends. Important observations relating to different states of droplet during its combustion are as follows:

(1) Jet A-1 and pure ethanol burned smoothly whereas all blends of ethanol/Jet A-1 showed disruptive nature. Apart from puffing and micro-explosion, abrupt explosion was also witnessed in the blends.

(2) The abrupt explosion and puffing are the dominant phenomena during the combustion of E10 droplets while micro-explosion is the most common event in droplets with a higher proportion of volatile component (E30 and E50).

(3) The most probable NOD for all the ethanol/Jet A-1 blends is observed to be nearly similar.

(4) The rate of bubble growth is almost the same for E30 and E50 droplets. However, bubble growth rate of ethanol blends is higher compared to that of butanol blends.

(5) As the proportion of ethanol is increased in the blends, the probability of micro-explosion increases while the droplet lifetime decreases.

(6) Puffing and abrupt explosion produce secondary droplets of smaller size as compared to those by micro-explosion. The SMD of the secondary droplets is around 20% of the initial droplet diameter.

(7) Both abrupt explosion and micro-explosion create large-scale flame distortion. The secondary droplets generated from abrupt explosion undergo faster evaporation compared to that of micro-explosion.

(8) The secondary droplets may undergo further fragmentation leading to the ejection of fine droplets. This feature is quite beneficial in the perspective of spray atomization.

## Acknowledgements

The authors are grateful to Indian Institute of Technology Kharagpur for providing financial support through Institute Scheme for Innovative Research and Development (ISIRD) program.